\documentclass[twocolumn,floatfix,prb,aps,showpacs,superscriptaddress,longbibliography]{revtex4-2}
\usepackage{amsmath, amssymb}

\usepackage{graphicx,amsmath,amssymb,color}
\usepackage{nicefrac}
\usepackage[titletoc,title]{appendix}
\usepackage{amsmath}
\usepackage{subfigure}
\usepackage{xcolor}
\usepackage{color}
\usepackage[colorlinks,bookmarks=true,citecolor=blue,linkcolor=red,urlcolor=blue]{hyperref}
\usepackage[colorlinks,bookmarks=true,citecolor=blue,linkcolor=red,urlcolor=blue]{hyperref}
\usepackage{titlesec}

\makeatletter
\def\@fnsymbol#1{%
  \ifcase#1\relax % Case 0 (unused)
  \or \ensuremath{\dagger}% Symbol 2: For the first \altaffiliation
  \or \ensuremath{*}% Symbol 3: For the second \altaffiliation
  \or \ensuremath{\mathsection}% Symbol 5
  \or \ensuremath{\mathparagraph}% Symbol 6
  \else\@ctrerr\fi% Fallback for more footnotes
}
\makeatother

\begin{document}

\title{Coexistence of static order and spin dynamics in an $S=5/2$ frustrated triangular antiferromagnet}

\author{U. Jena}
\thanks{equal contribution}
    \affiliation{
    Department of Physics, Indian Institute of Technology Madras, Chennai, 600036, India}
    
    \author{B. Sana}
    \thanks{equal contribution}
\affiliation{
Department of Physics, Indian Institute of Technology Madras, Chennai, 600036, India}
\author{Satish Kumar}
\affiliation{
Department of Physics, Indian Institute of Technology Madras, Chennai, 600036, India}

   \author{M. Pregelj}
\affiliation{Jo\v{z}ef Stefan Institute, Jamova c. 39, 1000 Ljubljana, Slovenia}
 \affiliation{Faculty of Mathematics and Physics, University of Ljubljana, Jadranska ulica 19, 1000 Ljubljana, Slovenia}
 \author{A. Bandyopadhyay}
 \affiliation{ISIS Neutron and Muon Source, Rutherford Appleton Laboratory,
 Harwell Campus, Didcot, OX11 0QX, United Kingdom}
 \affiliation{Department of Physics, R. B. College (A Constituent Unit of Lalit Narayan Mithila University, Darbhanga), Dalsingsarai, Samastipur, Bihar 848114, India}
 \author{P. Manuel}
 \affiliation{ISIS Neutron and Muon Source, Rutherford Appleton Laboratory,
 Harwell Campus, Didcot, OX11 0QX, United Kingdom}
 \author{J. S. Lord}
 \affiliation{ISIS Neutron and Muon Source, Rutherford Appleton Laboratory,
 Harwell Campus, Didcot, OX11 0QX, United Kingdom}
  \author{D. T. Adroja}
 \affiliation{ISIS Neutron and Muon Source, Rutherford Appleton Laboratory,
 Harwell Campus, Didcot, OX11 0QX, United Kingdom}
\author{P. Khuntia}
\email{pkhuntia@iitm.ac.in}
\affiliation{
Department of Physics, Indian Institute of Technology Madras, Chennai, 600036, India}
\affiliation{
Quantum Centre of Excellence for Diamond and Emergent Materials,
Indian Institute of Technology Madras, Chennai, 600036, India}
	\date{\today}
  % \thanks{equal contribution}

	\begin{abstract}

Frustrated triangular lattice antiferromagnets in the classical high-spin limit provide a paradigmatic setting in which the interplay of competing exchange interactions, anisotropy, and collective degrees of freedom can lead to unconventional low-energy excitations, anomalous criticality, and persistent dynamical responses. 
Here, we present comprehensive thermodynamic, $\mu$SR, and neutron diffraction experiments, along with first-principles calculations on a triangular lattice antiferromagnet, MnSnB$_2$O$_6$, where Mn$^{2+}$ ($S=5/2$) moments form a nearly perfect 2D triangular network without any anti-site disorder. The Curie–Weiss fit to the magnetic susceptibility yields a moderate Curie–Weiss temperature of $-12$ K, indicating the dominant antiferromagnetic interactions between Mn$^{2+}$ moments, which is supported by first-principles calculations.
Specific heat measurements reveal the onset of long-range magnetic order at $T_\text{N}\approx 1$ K, which is ascribed to intraplane exchange interactions.
Specific heat exhibits pronounced short-range correlations above $T_{\rm N}$ and an unconventional power-law behavior $C\propto T^{1.37}$ deep in the ordered state, suggesting the presence of non-trivial low-energy excitations.
%The temperature evolution of the order parameter down to 50 mK in neutron diffraction suggests that the ordered state is a classical 3D Ising-like antiferromagnet.
%Field-dependent specific heat measurements down to 50 mK reveal a gradual suppression of $T_\text{N}$ with increasing magnetic field, and $ T_\text {N} $ eventually disappears above an applied field of approximately 5 T. 
Zero-field $\mu$SR experiments down to 50 mK confirm the presence of magnetic ordering at 1 K, in agreement with thermodynamic and neutron diffraction experiments. The $\mu$SR measurements detect persistent spin dynamics coexisting with static magnetic order.
The temperature evolution of the order parameter down to 50 mK in neutron diffraction suggests that
the ordered state is consistent with a 3D Ising-like antiferromagnet.
%while specific heat exhibits pronounced short-range correlations above $T_\text{N}$ and an unconventional power-law behavior $C\propto T^{1.37}$ deep in the ordered state. 
The physics of this frustrated triangular lattice is governed by competing exchange interactions and exotic low-energy excitations. This family of archetypal frustrated magnets offers a promising venue for the experimental realization of emergent phenomena.
	
	\end{abstract}
	
	\maketitle

Frustrated magnets, where frustration-induced quantum fluctuations, a highly degenerate ground-state manifold, and interplay between collective degrees of freedom are operative, can host quantum states with exotic low-energy excitations, offering a viable ground to address some of the pressing questions in contemporary condensed matter~\cite{balents2010spin,khatua2023experimental,khuntia2020gapless,PhysRevLett.116.107203,khuntia2019novel}. In this context, geometrically frustrated triangular-lattice antiferromagnets (TLAFs) provide a prototypical platform for the experimental realization of diverse many-body quantum phases. The canonical nearest-neighbor Heisenberg antiferromagnet on the triangular lattice exhibits $120^\text{o}$ long-range magnetic order~\cite{PhysRevLett.60.2531}. However, frustrated TLAFs with next-nearest neighbor interactions on the $S=1/2$ system induces strong quantum fluctuations that can suppress the ordered moment and give rise to an array of unconventional phenomena, including quantum spin liquids (QSLs), field-induced spin-nematic states, and spin-density wave phases~\cite{khatua2023experimental, balents2010spin,PhysRevLett.129.087201,takagi1995new,PhysRevB.92.180411}. The TLAFs based on organic compounds such as $\kappa$-(BEDT-TTF)$_2$Cu$_2$(CN)$_3$~\cite{PhysRevLett.91.107001} and EtMe$_3$Sb[Pd(dmit)$_2$]$_2$~\cite{yamashita2010highly,PhysRevLett.123.247204} are some of the potential candidate QSL materials on the $S=1/2$ triangular lattice.
\begin{figure*}[htpb]
\begin{center}
%\includegraphics[height=3.5cm
%\textwidth]{}
\includegraphics[width=1\textwidth,height=0.3\textwidth]{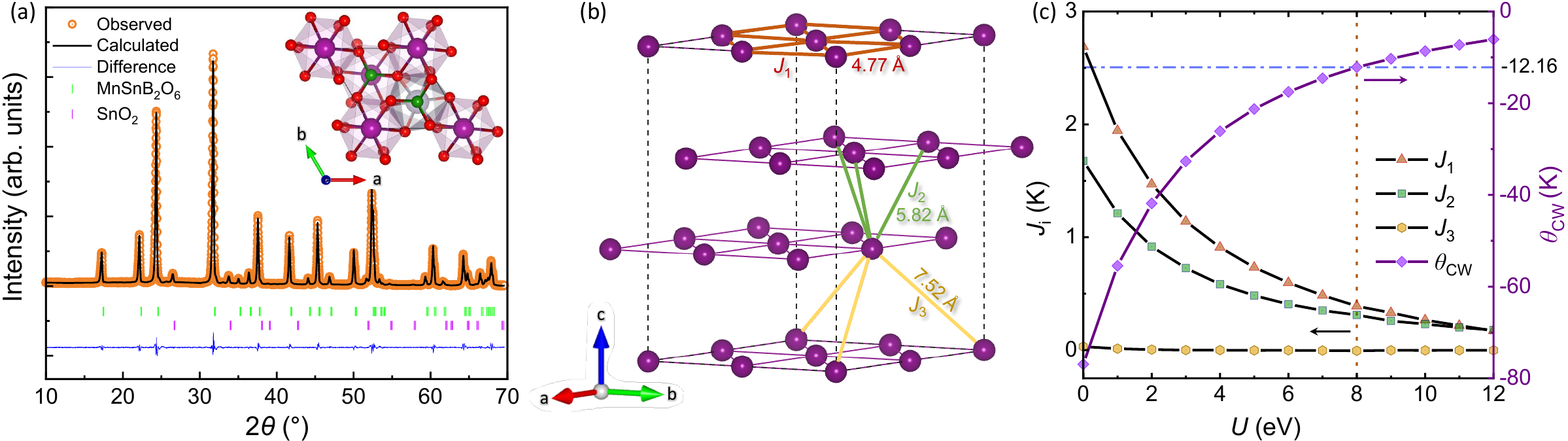}
  \caption{
  (a) The Rietveld refinement of the powder X-ray diffraction pattern of MnSnB$_2$O$_6$ taken at room temperature. A two-phase refinement is carried out due to the presence of unavoidable non-magnetic SnO$_2$ impurity phase (3.3\%), which does not affect the magnetic properties of the system. Inset shows a structural view of the $\text{MnO}_6$ octahedral layers observed along the $c$-axis.
  (b) Triangular lattice of Mn$^{2+} (S=5/2)$ moments in the crystallographic $ab$-plane with dominant exchange interaction pathways $J_1, J_2$ and $J_3$.
  (c) Dependence of characteristic magnetic parameters on the electronic correlation ($U$). The blue dashed-dotted line indicates the theoretical $\theta_{\text{CW}}$ value corresponding to the experimentally determined $\theta_{\text{CW}}$, obtained for $U = 8$ eV.
  }
  \vspace{-20pt}
\label{CrystalStructure}
\end{center}
\end{figure*}
Further, rare-earth (RE) triangular antiferromagnets provide a promising route for investigating unconventional magnetism, primarily driven by exchange anisotropy arising from the complex interplay between spin-orbit coupling (SOC) and crystal-electric field (CEF) effects. Specifically, RE magnets with an odd number of electrons in the $4f$ shell realize a $J_\text{eff}=1/2$ Kramers doublet ground state at low temperatures~\cite{PhysRevLett.115.097202,PhysRevB.106.104408, PhysRevB.111.094408}. As a result, the intrinsic quantum fluctuations suppress conventional long-range order, while a CEF-driven anisotropy yields to Ising-like or XY-like interactions~\cite{PhysRevX.9.021017,PhysRevResearch.3.043202,arh2022ising}. A recent study on the triangular heptatantalate family RETa$_7$O$_{19}$ revealed distinct exotic states, driven by the Ising-like nearest-neighbor spin correlations, including QSL~\cite{arh2022ising} and a spin Pomeranchuk effect~\cite{jaksetivc2026melting}, in which magnetic order melts upon cooling due to entropic and quantum effects. 
%In contrast to Kramers' ions, non-Kramer rare-earth ions can also host exotic magnetic behavior in frustrated triangular lattices, and one such prototypical example is TbInO$_3$, where experimental probes reveal a spin excitation continuum and short-range correlations characteristic of a two-dimensional spin liquid~\cite{clark2019two,jung2023unconventional}.

While TLAFs on quantum spin-$1/2$ are prime contenders for QSL physics, recent studies demonstrate that higher-spin systems are equally promising for realizing nontrivial many-body states~\cite{PhysRevB.104.184402}. In the ideal Heisenberg limit, the classical ground state of a TLAF exhibits an extensively degenerate manifold associated with the three-sublattice 120$^\circ$ order~\cite{villain1980order,collins1997review}. In the $J_1-J_2$ model, tuning the ratio of nearest to next-nearest neighbor exchanges leads to a rich phase diagram~\cite{PhysRevB.84.214418,PhysRevB.94.224413,struck2011quantum,PhysRevLett.102.237204,PhysRevLett.134.196702}. For example, in a classical TLAF with $J_1-J_2$ Heisenberg antiferromagnet, the conventional \(120^{\circ }\) Néel state is destabilized at \(J_2/J_1 = \frac{1}{8}\) and undergoes a first-order phase transition to a commensurate collinear state and subsequently evolves into an incommensurate spiral state with increasing \(J_2/J_1\)~\cite{PhysRevB.46.11137}. An applied magnetic field further enriches the phase diagram of classical TLAF, stabilizing exotic magnetic states such as Y, up-up-down ($uud$), and canted V phases, including the characteristic one-third magnetization plateau ~\cite{kawamura1984spin,gvozdikova2011magnetic}. An additional layer of richness is evident from the coupling between spin and charge degrees of freedom in TLAFs. In these systems, noncollinear Y and 120$^\circ$ phases give rise to multiferroicity~\cite{PhysRevLett.98.267205,PhysRevLett.108.237201}, linking magnetic order with spontaneous electric polarization, while ferroelectricity has also been observed in collinear $uud$ and high-field oblique phases in triangular-lattice antiferromagnets such as Ba$_3$NiNb$_2$O$_9$ and Ba$_3$CoNb$_2$O$_9$~\cite{PhysRevLett.109.257205,PhysRevB.89.104420}. The experimental realization of the 1/3 magnetic plateau has been reported in several classical Heisenberg TLAFs such as RbFe(MoO$_4$)$_2$ and Ba$_3$MnSb$_2$O$_9$~\cite{PhysRevLett.98.267205,tian2014susceptibility}.
While these states are well established in classical Heisenberg TLAFs, competing exchange interactions and spin anisotropy can substantially modify the magnetic phase diagram by lifting accidental degeneracies and stabilizing exotic magnetic phases~\cite{PhysRevB.84.214418}.

 Beyond static magnetic ordering, TLAFs in the symmetry-broken phase exhibit strongly renormalized spin excitations arising from competing interactions and magnetic anisotropies~\cite{PhysRevLett.133.096703}. For instance, below $T_{\rm N}$, TLAF FeI$_2$ with easy-axis single-ion anisotropy demonstrates that strong magnon–magnon interactions can give rise to multi-magnon bound states and hybridized excitation spectra beyond the conventional spin-wave description~\cite{PhysRevLett.127.267201}. The interplay between in-plane and easy-axis anisotropies in h-Y(Mn,Al)O$_3$ TLAF gives rise to exotic topological behavior, such as the impurity-driven extended spin textures which are defined as the spatially varying arrangement of magnetic moments~\cite{park2021spin}. %Importantly, this in-plane nature of the spin texture is also key to stabilizing certain gapped magnon modes. In the presence of Al$^{3+}$ impurities, it generates extended spin textures that fundamentally alter spin dynamics and offer a new paradigm for disorder in correlated systems. 
 Similarly, strong in-plane anisotropy in frustrated TLAF Ba$_2$La$_2$CoTe$_2$O$_{12}$ tunes the spin dynamics by flattening the magnon dispersion and generating higher-order van Hove singularities in the ordered state~\cite{park2024anomalous}.   
% Further complexity appears in the metallic frustrated triangular-lattice, where the interplay between geometric frustration and electronic degrees of freedom stabilizes topologically nontrivial skyrmions, as observed in Gd$_2$PdSi$_3$~\cite{kurumaji2019skyrmion}. 
 Thus, classical TLAFs with reduced quantum fluctuations offer a rich landscape for the experimental realization of non-trivial ground states with exotic collective low-energy excitations driven by frustration, competing interactions, and anisotropies.
 However, classical TLAFs are relatively less explored compared to its quantum counterparts due to the scarcity of ideal candidate materials. 
 In this context, the search for promising disorder-free $S > 1/2$ TLAFs is of significant interest because these materials can stabilize complex many-body phenomena with exotic collective excitations and offer a compelling platform for distinguishing the effects of geometric frustration from those driven by strong quantum fluctuations in their $S = 1/2$ counterparts~\cite{bx3h-4g62,nakatsuji2005spin,lee2002emergent}.

\begin{figure*}[ht]
%\centering
%\includegraphics[height=3.5cm
%\textwidth]{}
\includegraphics[width=1\textwidth]{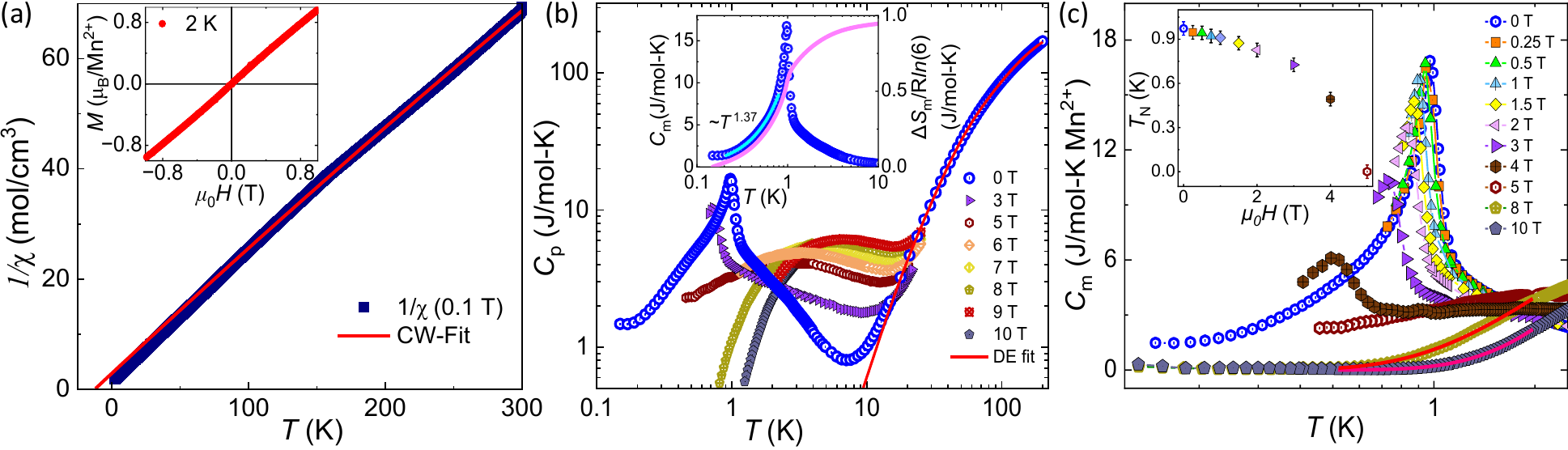}
  \caption{(a) Temperature dependence of the \textit{dc} inverse magnetic susceptibility measured under an applied field of 0.1 T. The inset shows a five-quadrant magnetization (\textit{M–H}) curve recorded at 2 K, exhibiting no hysteresis. (b) Temperature dependence of the specific heat $C_\text{p}$(\textit{T}), measured in zero-field and in applied magnetic fields. The zero-field magnetic specific heat, obtained after subtracting lattice contributions following the Debye-Einstein model, and the corresponding entropy are depicted in the inset. (c) Evolution of the specific heat anomaly under applied magnetic fields. Red solid lines depict the fit to $C_\text{m}\sim   e^{-\Delta/k_\text{B}T}$ for 8 T and 10 T. Inset: \(H\)–\(T\) phase diagram constructed from the Néel temperature \(T_{\text{N}}\) as a function of the applied magnetic field \(H\).}
  \vspace{-15pt}
\label{Fig.2 MSBO}
\end{figure*}

Here, we present our investigations on a structurally perfect frustrated magnet, MnSnB$_2$O$_6$ (henceforth MSBO), in which Mn$^{2+}$ ($S = 5/2$) ions form an equilateral triangular lattice stacked along the crystallographic $c$-axis. The specific heat measurements, complemented by neutron diffraction and $\mu$SR probes, establish the onset of long-range antiferromagnetic order at $T_\mathrm{N}\sim1$ K arises inter-layer interactions, substantiated by first-principles calculations. The specific heat shows an unconventional power-law behavior below \(T_{\text{N}}\), which suggests a softening of the spin-wave stiffness due to nearly degenerate states arising from competing interactions.
The broad anomaly in the zero-field specific heat above $T_{\rm N}$ indicates the development of short-range spin correlations, which further supported by the fact that only $\sim 90\%$ of the total magnetic entropy expected for an $S=5/2$ system is recovered at $T_{\rm N}$.
Neutron diffraction indicates the presence of in-plane exchange anisotropy and reveals 3D Ising-like universality. Furthermore, $\mu$SR relaxation measurements reveal the coexistence of static and fluctuating moments in the magnetically ordered state down to 50 mK, highlighting the effect of geometric frustration and anisotropy leading to an unconventional ground state in this $S=5/2$ frustrated magnet.

The polycrystalline sample of MSBO was synthesized following a solid-state reaction route. MSBO crystallizes in the trigonal space group 
$R\bar{3}$~\cite{cooper1994crystal}, without any structural disorder within the experimental (XRD+NPD) resolution [see Fig.~\ref{CrystalStructure}(a)]. The refined crystallographic parameters are summarized in Supplementary Material (SM) Table~1~\cite{supplement}. 
The magnetic sublattice is formed by Mn$^{2+}$ ions, which, due to their half-filled 3$d^5$ electronic configuration, can be described as nearly ideal Heisenberg spins with spin quantum number $S = 5/2$ and a quenched orbital angular momentum. Within the crystallographic $ab$-plane, the Mn$^{2+}$ ions occupy equivalent positions on an equilateral triangular network, thus providing a natural platform for studying geometrical frustration as shown in Fig.~\ref{CrystalStructure}(b). The estimated exchange coupling constant $J_i$\,(K) as a function of electronic correlation $U$\,(eV) depends strongly on the specific coordination pathways [see Fig.~\ref{CrystalStructure}(c)]. Specifically, the intra-layer interactions ($J_1$) proceed via the Mn-O-B-O-Mn exchange path, while inter-layer interactions ($J_2$ and $J_3$) share a common Mn-O-Sn-O-Mn path but differ in bond angles~\cite{supplement}.
%Schematic representations of these pathways are discussed in detail.

%
\begin{figure*}[!]
\centering
\includegraphics[width=\textwidth]{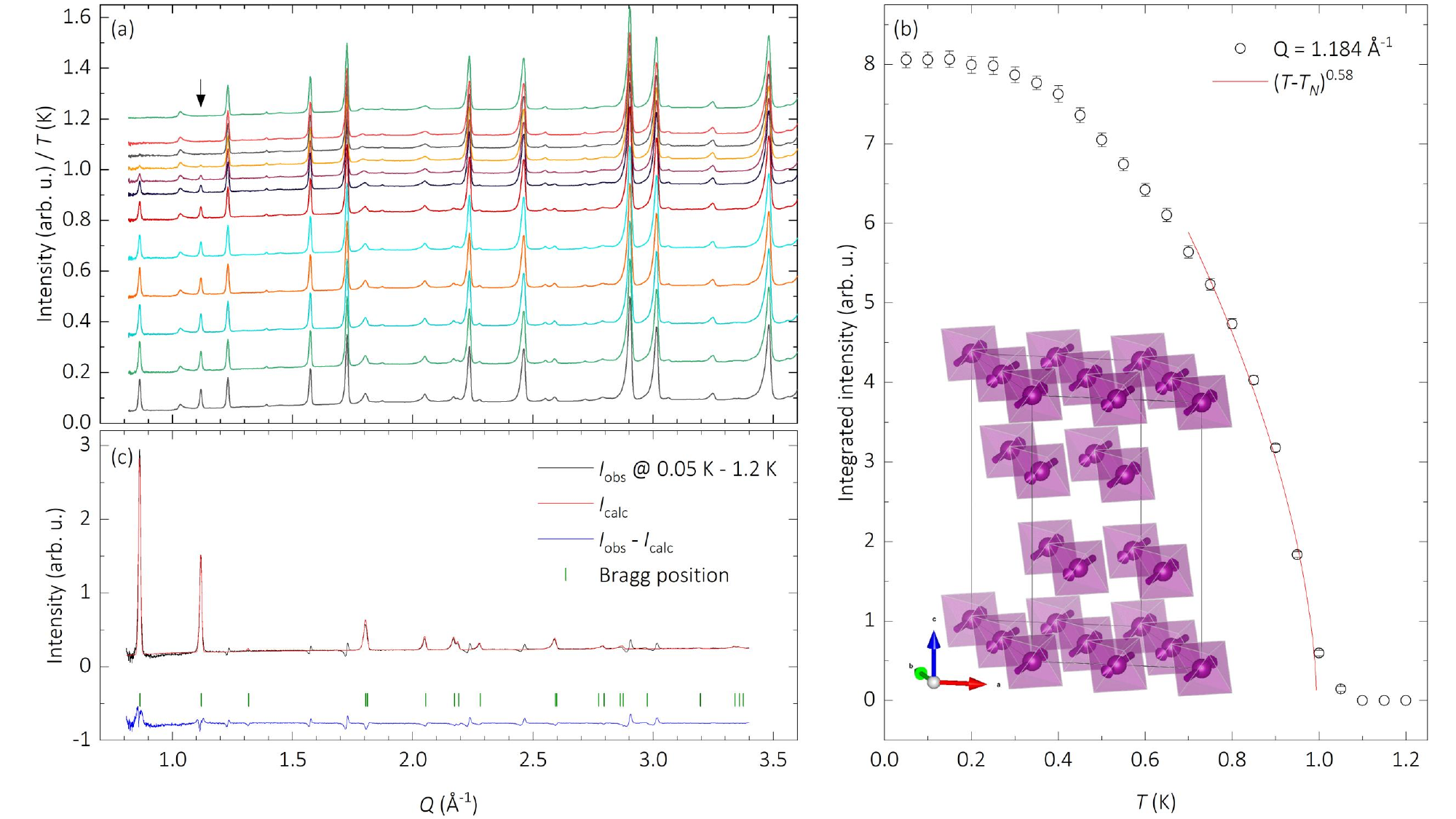}
\caption{(a) Neutron powder diffraction patterns for MnSnB$_2$O$_6$  measured between 0.05 and 1.2\,K. The intensity scale also corresponds to the temperature at which diffraction patterns were measured. The arrow denotes the position of a pronounced magnetic reflection. (b) The derived temperature dependence of the marked magnetic-peak intensity. The inset shows the refined magnetic ground state. (c) The difference between the data measured at 0.05\,K and 1.2\,K compared with the calculated diffraction pattern and corresponding residual, signifying a very good agreement.}
\vspace{-15pt}
\label{fig-all}
\end{figure*}

The temperature dependence of the magnetic susceptibility, $\chi(T)$, was measured down to 2~K under an applied field of 0.1~T [Fig.~\ref{Fig.2 MSBO}(a)]. A Curie--Weiss analysis of magnetic susceptibility,  
$
\chi(T) = \chi_0 + \frac{C}{(T-\theta_{\rm CW})},
$ 
yields a characteristic Curie--Weiss temperature $\theta_{\rm CW} = -12.4(2)$~K, and Curie constant $C = 4.26(5)~{\rm cm^3\,K\,mol^{-1}}$. The temperature-independent term $\chi_0 = 7.70\times10^{-4}~{\rm cm^3mol^{-1}}$ includes both core diamagnetic and Van Vleck contributions, where the estimated core diamagnetism $\chi_{\rm dia} = -1.02\times10^{-4}~{\rm cm^3mol^{-1}}$ and Van Vleck susceptibility $\chi_{\rm VV} \approx 8.72\times10^{-4}~{\rm cm^3mol^{-1}}$ [more details are given in SM~\cite{supplement}]. The effective magnetic moment is estimated to be $\mu_{\rm eff} = \sqrt{8C}\,\mu_B = 5.83~\mu_B$, in close agreement with the spin-only value of $5.92~\mu_B$ expected for high-spin Mn$^{2+}$ ($3d^5$, $S=5/2$, $g\approx 2$) ions. The negative value of $\theta_\text{CW} \approx -12$ K indicates predominantly antiferromagnetic exchange interactions between Mn$^{2+}$ moments. 
The five-quadrant $M(H)$ loop at 2 K shows no hysteresis or remanence, indicating the absence of a sizable ferromagnetic component within the experimental resolution [see the inset of Fig.~\ref{Fig.2 MSBO}(a)].

Specific heat measurements provide a direct probe of phase transition, spin correlations, and the presence of exotic low-energy magnetic excitations in frustrated magnets. The zero-field specific heat, shown in Fig.~\ref{Fig.2 MSBO}(b), reveals a sharp $\lambda$-like anomaly at $T_{\rm N} \approx 1$~K, signaling the onset of long-range antiferromagnetic order of the Mn$^{2+}$ moments. 
In the absence of non-magnetic analog, magnetic specific heat $C_\text{m}(T)$ was obtained by subtracting the lattice contribution, estimated using a Debye–Einstein function, from the total specific heat, $C_\text{p}(T)$, [see SM~\cite{supplement}].
The magnetic entropy release [see inset of Fig.~\ref{Fig.2 MSBO}(b)], obtained by integrating $C_{\mathrm{m}}(T)/T$, is approximately 90\% of $R\ln(6)$ at $T_{\rm N} =1$ K, which is lower than that expected for $S=5/2$ system, indicates the presence of short-range correlations, consistent with the broad maximum observed above 1 K in zero-field specific heat [Fig.~\ref{Fig.2 MSBO}(b)]. The zero-field magnetic specific heat exhibits a power-law dependence $T^{1.37}$ much below the magnetic transition temperature. Such unconventional power-law behavior arises from competing interactions that result in degenerate states, softening the spin-wave stiffness over a continuum of wave vectors~\cite{bergman2007order,plumb2019continuum}.

%In geometrically frustrated systems, competing interactions create a manifold of nearly degenerate states that gives rise to a proliferation of low-energy fluctuations beyond symmetry-broken magnon modes, and theoretical studies show that this degeneracy can produce extended soft-mode manifolds in momentum space, where the spin-wave stiffness vanishes over a continuum of wave vectors~\cite{bergman2007order,plumb2019continuum}. 

 As shown in Fig.~\ref{Fig.2 MSBO}(c), the sharp anomaly at $T_\text{N}$ progressively shifts to lower temperatures with increasing field strength upto 4 T, reflecting the suppression of antiferromagnetic order [see the inset of Fig.~\ref{Fig.2 MSBO}(c)]. 
 In an isotropic unfrustrated magnet, the ordered state is expected to be suppressed when the Zeeman energy ($g\mu_\text{B}H\sqrt{S(S+1)}$) becomes comparable to the characteristic exchange energy scale ($k_\text{B}T_\text{N}$)~\cite{PhysRevLett.84.2957}. This yields an estimate of the critical field $H_\text{C}(0)=k_BT_\text{N}/g\mu_\text{B}\sqrt{S(S+1)}\approx 0.25 $ T. The presence of anisotropy and competing exchange interaction likely plays a significant role in reinforcing the ordered phase against stronger external magnetic fields ($>0.25$ T). As depicted in Fig.~\ref{Fig.2 MSBO}(b), broad maxima appear in $C_{\rm p}(T)$, for applied fields above 5 T, indicating a shift to higher temperatures with increasing magnetic field up to 10 T. With the application of high magnetic fields ($\mu_0H=8$ T and 10 T), the low-temperature specific heat exhibits a rapid suppression towards zero as shown in Fig.~\ref{Fig.2 MSBO}(c). The data are well described by a thermally activated behavior $C_\text{m}\sim   e^{-\Delta/k_\text{B}T}$, yielding field dependent energy gaps that increase with field, reflecting 
 the emergence of a field-polarized state, as independently supported by the $M$ vs $H$ curve [Fig.\,S2(b)~\cite{supplement}], which saturates above 7 T at 2 K. In a spin polarized regime, the low-energy excitations correspond to single spin-flip processes~\cite{zhitomirsky2005high,PhysRevLett.91.177601}, which are described by the Zeeman splitting relative to the saturation field, $\Delta\sim g\mu_B(H-H_\text{sat})$. 
 %The H–T phase diagram [inset of Fig.~\ref{Fig.2 MSBO}(c)] was obtained from the field dependence of $T_\text{N}$ fitted with Curie-Bloch expression~\cite{marfoua2024ultra}: $H(T)=H(0)\left[1-\left(T/T_\text{N}\right)^{\alpha}\right]^{\beta'}$ in the low temperature regime, where $H(0)=5$ T and a critical exponent $\beta'=0.4(8)$, which is consistent with 3D magnetic ordering~\cite{pelissetto2002critical}. The low-temperature critical exponent $\alpha=1.47(1)$, which is close to the expected value of $\alpha=\frac{1}{2}$ for two-dimensional half-integer spin systems~\cite{kobler2002temperature}, consistent with the quasi-2D spin 5/2 triangular magnetic lattice of MSBO.

\begin{figure*}[htbp]
    \centering
    \includegraphics[width=1\linewidth]{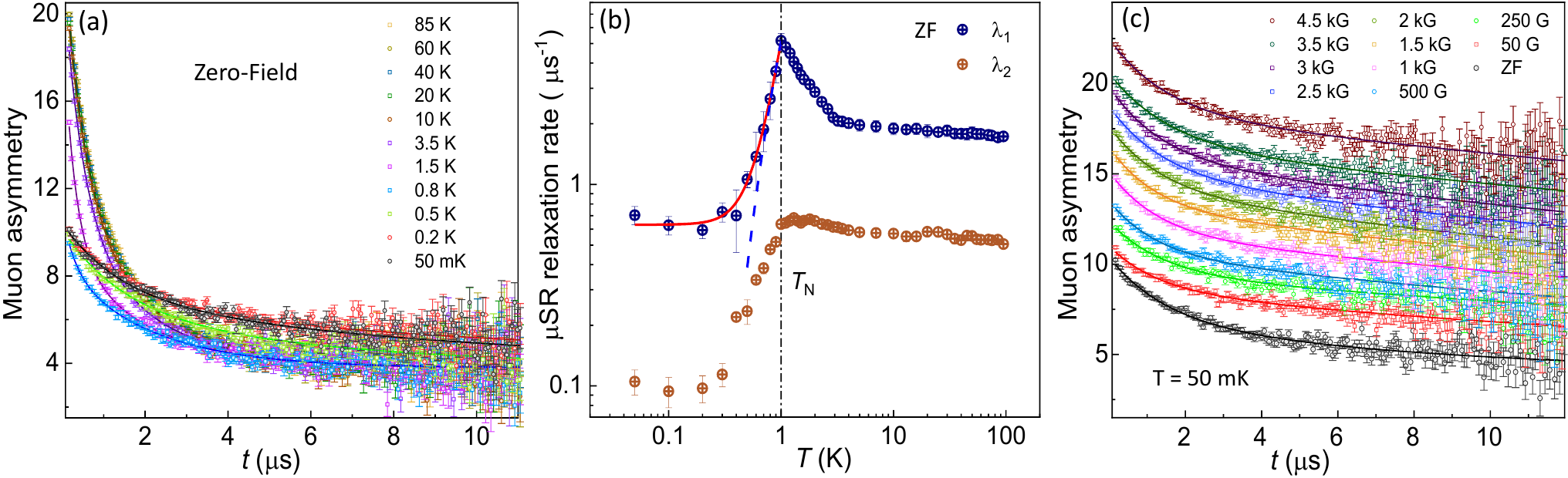}
    \caption{(a) ZF-$\mu $SR asymmetry spectra measured at representative temperatures. (b) Temperature dependence of the fast ($\lambda_1$) and slow ($\lambda_2$) $\mu $SR relaxation rates in zero-field. (c) Longitudinal-field (LF) $\mu $SR asymmetry spectra recorded at selected LF values at T = 50~mK. All corresponding fits are discussed in the text.}
    \vspace{-10pt}
    \label{fig:muSR}
\end{figure*}

\begin{table}[b]
 \centering
 \caption{The irreducible representation (IRR) $\Gamma_1$ and corresponding basis vectors $\psi_1^j$ for the space group $R$-3 appearing in the magnetic representation with $\mathbf{k}=(1.5~0~0)$ for the Mn magnetic site (0.0,~0.0,~0.0), with magnetic unit cell corresponding to the $R$-1 space group. We note that representation analysis was performed using BasIreps program incorporated in the FullProf Suite \cite{rodriguez1990fullprof}.}
  \begin{tabular}{ c c c c  }
    \hline \hline
      ~~~~~~~vector ~~~~~~~~&~~~~~~~$m_a$ ~~~~~~& ~~~~~~~$m_b$ ~~~~~~~& ~~~~~$m_c$   \\
\hline
     $\psi_1^1$ &  1  &  0  &  0  \\
     $\psi_1^2$ &  0  &  1  &  0  \\
     $\psi_1^3$ &  0  &  0  &  1  \\
   \hline \hline
 \end{tabular}
 \label{tab-irreps}
\end{table}

To gain insights into the antiferromagnetically ordered state and the corresponding order parameter in this frustrated magnet, we carried out neutron diffraction studies down to 50 mK. The neutron diffraction patterns for MSBO were recorded between 0.05 and 100 K, and the low temperature data are shown in Fig.~\ref{fig-all}(a).
The diffraction pattern measured at 1.2\,K can be reasonably well refined to the expected crystal structure, yielding an Rf-factor of 12.2.
Below $\sim$1\ K, additional Bragg reflections emerge [arrow in Fig.~\ref{fig-all}(a)], which we ascribe to the establishment of the long-range magnetic order.
The integrated intensity of the magnetic reflection at $Q$\,=\,1.184\,\AA$^{-1}$~exhibits a clear order-parameter-like response [Fig.~\ref{fig-all}(b)], and can be described by $I\sim(T-T_{\rm N})^{2\beta}$, where $T_\text{N}$\,=\,1.00(1)\,K is the magnetic transition temperature and $\beta$\,=0.29(2) is the critical exponent. It may be worth noting that the factor of 2 in the exponent accounts for the fact that the magnetic-reflection intensity is proportional to the square of the magnetic moment. The value of the critical exponent is significantly lower than that expected for conventional 3D Heisenberg ($\beta=0.36$)~\cite{PhysRevB.65.144520} and close to that expected for 3D  Ising ($\beta=0.32$) universality class~\cite{pelissetto2002critical}, indicating that the low temperature physics is governed by Ising-like magnetic anisotropy. The anisotropic antiferromagnetic structure further revealed by neutron diffraction suggests that magnetic anisotropy drives the critical behavior away from the isotropic 3D Heisenberg toward a 3D Ising-like universality class~\cite{PhysRevLett.128.015703,akutsu1981specific}.

To determine the magnetic structure, we first subtracted the paramagnetic signal from the base temperature data by subtracting the 1.2 K data from the 0.05 K data.
The resulting difference plot is shown in Fig.~\ref{fig-all}(c).
The position of the magnetic reflections can be described with the (1.5~0~0), (0~1.5~0), as well as (1.5~$-$1.5~0) magnetic wave vectors. 
The representation analysis based on any of these magnetic wave vectors yields the same one-dimensional irreducible representation (IRR), consisting of three basis vectors [see Table~\ref{tab-irreps}], while the magnetic unit cell corresponds to the $R$-1 magnetic space group. These basis vectors were then used as an input for the magnetic structure refinement of the difference pattern between the 0.05 K and 1.2 K data.
A reasonably high quality of the refinement (Rf-factor of 10.2) [Fig.~\ref{fig-all}(c)] yielded $\psi_1^1$\,=\,$4.58(1)$, $\psi_1^2$\,=\,$4.58(1)$, and $\psi_1^3$ is below the experimental sensitivity.
The corresponding ordered magnetic moment of the Mn ion is thus ($m_a$,\,$m_b$,\,$m_c$)\,=\,$[4.58(1),\,4.58(1),\,0]$\,$\mu_\text{B}$, where $m_i$ are magnetic moment components along the $a$, $b$ and $c$ crystallographic axes, yielding the ordered magnetic moment size of 4.58(1)\,$\mu_\text{B}$, i.e., in reasonable agreement with the  Mn$^{2+}$ $S$\,=\,5/2.
The corresponding magnetic structure, presented in the inset of Fig.~\ref{fig-all}(b), reveals a stripe magnetic structure with a predominantly antiferromagnetic ordering within the $ab$-plane.

%time evolution of ZF-$\mu $SR spectra at various temperatures for MSBO. The solid curves represent the fitting as a two-component exponential relaxation model. (b) Longitudinal-field $\mu$SR asymmetry spectra measured at a fixed temperature under various applied magnetic fields, fitted with the same model. (c) Temperature dependence of the extracted muon spin relaxation rates, $\lambda_1$ and $\lambda_2$, derived from the zero-field data in panel (a). (d) Magnetic field dependence of the relaxation rates, $\lambda_1$ and $\lambda_2$, obtained from the longitudinal-field measurements in panel (b).

To probe the static magnetism, phase transition, and low-energy spin dynamics in the magnetically ordered state, we performed zero-field (ZF) and longitudinal-field (LF) $\mu$SR measurements on MSBO down to 50 mK~[see Fig.~\ref{fig:muSR}(a,c)]. We first examined the temperature evolution of the ZF-$\mu$SR asymmetry across the magnetic transition in MSBO. A pronounced reduction in the initial asymmetry is observed upon cooling from 1.5 K to 0.8 K. This substantial loss of initial asymmetry reflects the onset of internal magnetic fields at the muon site associated with long-range order. The absence of oscillatory behavior in the ZF-$\mu$SR asymmetry in the ordered state can be attributed to the finite pulse width ($\sim 70$ ns) of the pulsed muon source (ISIS, UK), which limits the observation of the damping of the fast relaxing component at short times. The ZF spectra are well described by a two-component exponential relaxation function of the form $
A(t) = A_{bkg} + A_0\left[ f \exp(-\lambda_1t) + (1-f) \exp(-\lambda_2t)\right]$, where $\lambda_1$, $\lambda_2$ represent fast and slow relaxing components, respectively. Here $A_{0}$ denotes the initial asymmetry, $f= 0.66$ represents the relative weight fraction of the fast-relaxing component with respect to the slow relaxing part, and $A_{\mathrm{bkg}}$ accounts for the temperature-independent background contribution that was kept fixed throughout the measured temperature range. The temperature dependence of $A_0$  and $f$ is shown in the SM~\cite{supplement} [Fig. S3] and reflects the establishment of magnetic ordering

As shown in Fig.~\ref{fig:muSR}(b), the fast relaxation rate $\lambda_1$ exhibits a pronounced peak at $T_\text{N}\sim 1$ K,  signaling the slowing down of spin fluctuations and the onset of long-range magnetic ordering, which is in agreement with the specific heat and neutron diffraction results. The slow component $\lambda_2$ shows a similar $T$-dependence. In contrast to the conventional magnetic phase transition where the relaxation rate rapidly decreases below $T_\text{N}$, both relaxation rates remain sizeable down to 50 mK. In ZF $\mu$SR, the fast relaxation rate \(\lambda _{1}\) reaches a plateau at about 0.7 \(\mu\)s\({}^{-1}\) constant down to 50 mK. A similar behavior has been observed in other frustrated magnets, such as pyrochlore Gd\({}_{2}\)Sn\({}_{2}\)O\({}_{7}\), where the relaxation rate decreases by an order of magnitude from its maximum value at \(T_{\text{N}}\) and plateaus at 0.6 \(\mu\)s\({}^{-1}\) down to 20 mK~\cite{mcclarty2011calculation}. This behavior suggests the persistence of low-energy spin dynamics within the magnetically ordered state, a characteristic feature observed in several frustrated magnets~\cite{PhysRevB.99.214441}. Below $T_\text{N}$, the fast relaxation rate follows a power law $\lambda_1=a+bT^{n}$ [red solid line in Fig.~\ref{fig:muSR}(b)] with $n=3.4$ and the finite value of $a=0.7$ $\mu$s$^{-1}$ consistent with persistent dynamics scenario. For a 3D antiferromagnet with gapless spin wave excitations due to a two-magnon Raman scattering process, a $T^3$ dependence of the relaxation rate is expected, whereas higher-order magnon processes can lead to larger exponents~\cite{PhysRevLett.103.077207}. Considering a spin-wave spectrum gap of \(E_{g}\), \(\lambda _{1}\) was also fitted to an exponential form of \(T\exp(-E_g/T)\) [blue dashed line in Fig.~\ref{fig:muSR}(b)]; however, the poor fit quality suggests the absence of a sizable gap, and indicates the low energy excitations are mainly due to gapless modes~\cite{PhysRevLett.109.227202}.

The LF $\mu$SR asymmetry was measured at 50 mK over the field range 0–0.45 T [Fig.~\ref{fig:muSR}(c)] to shed insights into the spin dynamics. All LF-$\mu$SR asymmetry spectra were analyzed using the same fitting function employed for the ZF data, with the parameters $f_0$ and $A_\text{bkg}$ fixed to the values obtained from the corresponding fits. The incomplete recovery of the LF-$\mu$SR asymmetry spectra at 4.5 kG suggests the persistence of low-energy spin dynamics in the ordered state. In addition, both relaxation rates, $\lambda_1$ and $\lambda_2$, exhibit a monotonic decrease with increasing LF; however, suppression of muon relaxation remains incomplete even at the highest LF of 4.5 kG. 
%Rather, significant muon-spin relaxation persists at 4.5 kG [Fig.\,S3(a)~\cite{supplement}], as also reflected in the LF-dependent variations of the relaxation rates [Fig.\,S3(e)~\cite{supplement}] by the substantially finite values of both relaxation rates, indicating persistent spin dynamics. 
The incomplete muon decoupling in the external LF does not originate from the internal static fields due to the Mn local moments. Instead, the persistent muon-spin relaxation in the 50 mK LF-$\mu$SR spectra could be attributed to the presence of dynamic magnetic fluctuations in the AFM-ordered ground state, further adding credence to the ZF-$\mu$SR results.

Density functional theory calculations (PBE + $U=8$ eV on Mn 3$d$) confirm MSBO as an antiferromagnetic insulator with $E_g = 2.18$ eV, where correlations split Mn $d$ states into Hubbard bands. Exchange interactions, obtained from Noodleman’s broken-symmetry analysis of FM and AFM supercell energies [Fig.\,S4(a-d)~\cite{supplement}], yield competing energy scales $J_{1} = 0.389~\mathrm{K}$ (intra-layer, $4.77~\text{\AA}$),
$J_{2} = 0.311~\mathrm{K}$ (inter-layer, $5.82~\text{\AA}$, $\approx 0.8\,J_{1}$),
and $J_{3} = -0.006~\mathrm{K}$ ($7.52~\text{\AA}$, weakly ferromagnetic) [see Fig.~\ref{CrystalStructure}(c)].
%$J_1$ exchange coupling path is mediated through a Mn-O-B-O-Mn super-superexchange pathway with a Mn-Mn separation of 4.77 \AA, while $J_2$ and $J_3$ is mediated through a Mn-O-Sn-O-Mn with a Mn-Mn separation of 5.82 \AA and 7.52 \AA, respectively.
These exchange parameters yield the expected experimental value for CW temperature of $\theta_{\mathrm{CW}} \approx -12~\mathrm{K}$ following mean-field approximation, and the relatively strong $J_{2}$ interaction partially relieves the frustration, thereby stabilizing a collinear magnetic ground state~\cite{PhysRevB.46.11137}.

Neutron diffraction experiments reveal a 3D Ising-like critical behavior, and the slightly reduced value of the exponent compared to that expected for the 3D Ising ($\beta=0.32$) universality class suggests that critical spin fluctuations are not purely isotropic, likely due to residual anisotropy and/or quasi-2D spin correlations inherent to the frustrated triangular lattice~\cite{akutsu1981specific}. The refined magnetic moment components allow us to quantify the deviation from the ideal coplanar configuration. 
%Using the crystallographic metric appropriate for the trigonal lattice MSBO, the in-plane component of the ordered moment is given by $m_{ab}=\sqrt{m_\text{a}^2+m_\text{b}^2-m_\text{a}m_\text{b}}$, and the out-of-plane canting angle is estimated as $\theta=\tan^{-1}(m_\text{c}/m_\text{ab})\approx10^{\circ}$. Another TLAF, Ba$_3$MnNb$_2$O$_9$, exhibits similar out-of-plane canting with an angle of 18.7\({}^{\circ }\) and shows multiferroicity at low temperatures~\cite{PhysRevB.90.224402}. In an ideal Heisenberg triangular antiferromagnet, spins adopt a coplanar 120° configuration; thus, the observed canting reflects the presence of additional anisotropic interactions, which is consistent with the expected 3D Ising nature of MSBO~\cite{pelissetto2002critical}. 
 %The ordered magnetic moment of 4.65(9) $\mu_\text{B}$ is close to the expected classical value for Mn$^{2+}$ ($S=5/2$), suggesting weak quantum fluctuations and placing this system in the regime of classical frustrated magnet.
 For an ordered ground state in the large-$S$ limit, the O(3) nonlinear $\sigma$-model for a nearest-neighbor Heisenberg antiferromagnet gives spin-wave velocity $
c = \frac{2\sqrt{D} J_\text{eff} S d_n}{\hbar}$~\cite{auerbach2012interacting}, where $D$ is spin-lattice dimensionality. Using dominant $J_1$ and $J_2$, $J_\text{eff} = (z_1 J_1 + z_2 J_2)/(z_1 + z_2) = 0.35$ K ($z_1 = 6$, $z_2 = 6$) and Mn--Mn nearest-neighbor distance $d_n$, resulting $c \sim 1.9 \times 10^2$ m/s. For frustrated $S = 1$ triangular-lattice NiGa$_2$S$_4$, frustration renormalizes $c = 8.5 \times 10^2$ m/s, with quadratic low-$T$ specific heat~\cite{nakatsuji2005spin}. The relatively reduced value of $c$ in MSBO suggests an enhancement of the density of low-energy states, as evident from the $T^{1.37}$ power-law dependence of the zero-field specific heat below $T_\mathrm{N}$. It suggests renormalization of the spin-wave spectrum in the presence of nearly degenerate states arising from magnetic frustration. Furthermore, $\mu$SR experiments probe an unconventional ground state marked by the coexistence of static long-range antiferromagnetic order and persistent spin dynamics down to 50 mK with non-trivial low-energy excitations.
%Further, the $\mu$SR results reveal a magnetic ground state that deviates from conventional long-range ordered magnets. While the peak in the relaxation rate at $T_\text{N}\sim 1$ K signals the onset of magnetic order via critical slowing down, the persistence of sizeable relaxation rates down to 50 mK indicates that spin dynamics are not completely frozen in the ordered state. The incomplete decoupling of the $\mu$SR asymmetry under applied LFs, even up to 0.45 T, establishes the presence of strong dynamic spin fluctuation. 
This behavior closely parallels the persistent spin dynamics observed in other classical spin systems such as the triangular NaCrO$_2$~\cite{PhysRevLett.97.167203} and the spin-chain system FeTe$_2$O$_5$Br even in the ordered state~\cite{PhysRevLett.109.227202}. In FeTe$_2$O$_5$Br, amplitude-modulated magnetic order naturally accommodates the coexistence of ordered and fluctuating spin components. Such a scenario in the present case is most likely attributed to competing exchange interactions and the characteristic magnetic structure of  this frustrated antiferromagnet~\cite{PhysRevLett.109.227202}.

To summarize, we studied the magnetic ground state of a nearly perfect frustrated triangular lattice antiferromagnet MSBO through a combination of thermodynamic, $\mu$SR, neutron-diffraction, and first-principles investigations. DFT+U calculations indicate that MSBO is an antiferromagnetic insulator with a band gap of 2.18 eV. Neutron diffraction reveals a collinear magnetic structure in the crystallographic $ ab$-plane. The Ising-like character of the magnetic ordering evident from the order-parameter analysis and refinement of magnetic structure, suggests the presence of magnetic anisotropy. The magnetic specific heat suggests the presence of short range spin correlations above $T_\text{N}$ and non-trivial low-energy excitations below $T_\text{N}$. In addition, $\mu$SR experiments reveal the coexistence of static magnetic order and persistent spin fluctuations, highlights the role of competing exchange interactions and geometric frustration in stabilizing unconventional low temperature physics in this classical spin systems. From a theoretical perspective, the system offers a promising platform for investigating how weak competing exchange  interactions and the stacking triangular geometry affect the crossover between classical 3D Ising-like behavior and fluctuation-driven regimes. These directions may provide broader insight into the emergence of unconventional dynamics in moderately frustrated three-dimensional classical
magnets.

\noindent\hspace{1em} \textit{Acknowledgements.}
P.K. acknowledges the funding by the Anusandhan National Research Foundation (ANRF), Department of Science and Technology, India through Research Grants. M.P. acknowledges the funding by the Slovenian Research Agency (program No. P1-0125). D.T.A. would like to thank EPSRC UK for the funding (Grant No. EP/W00562X/1).

\vspace{0.75em} \textit{Data availability.}
The data that support the findings of the current study are available from the corresponding author upon reasonable request.

\bibliography{MSBO}
%\bibliographystyle{apsrev4-2}
%\printbibliography

\end{document}